Eversion Buckling of Toroidal Shells: Mechanism, Bifurcation, and Applications in Energy Absorption

Junjie Liu[a,1,*], Aijie Tang[a,1], Mingchao Liu[b], Xiaoding Wei[c], Qingsheng Yang[a,*]

Address:

[a] Department of Mechanics, Beijing University of Technology, Beijing 100124, China.

[b] Department of Mechanical Engineering, University of Birmingham, Birmingham B15 2TT, UK

[c] State Key Laboratory for Turbulence and Complex System, HEDPS, Center for Applied Physics and Technology, Department of Mechanics, School of Mechanics and Engineering Science, Peking University, Beijing 100871, China

[1] These authors contributed equally to this work.

[*] Corresponding authors,

E-mails: liujunjie@bjut.edu.cn; qsyang@bjut.edu.cn

**Abstract:**

Thin shells can store and release elastic energy through large geometric transformations, but their post-deformation stability is often highly sensitive to geometry and loading direction. Here, we study the eversion of open toroidal shells generated by revolving a meridional arc about an external axis and identify a geometry-controlled instability, which we define as eversion buckling. After eversion, the shell may either remain in an axisymmetric everted state or spontaneously collapse into a non-axisymmetric configuration. We show that this transition is governed by the competition between bending and membrane energies. A scaling analysis leads to a dimensionless geometric parameter that predicts the onset of eversion buckling, namely, the spontaneous loss of stability of the axisymmetric everted configuration after unloading. Finite element simulations and experiments confirm that this parameter collapses the stability data for shells with different geometries, including circular and semi-elliptical generating curves. The loss of axisymmetric stability is consistent with a pitchfork-like symmetry-breaking transition, in which the collapsed state has no preferred in-plane direction because of the axisymmetry of the everted shell. When the axisymmetric everted configuration is stable, a finite perturbation can trigger rapid snap-through toward a lower-energy collapsed equilibrium, accompanied by substantial volume contraction and a triggering response that is relatively insensitive to boundary constraints. We further show that assemblies of these shells behave as granular energy-absorbing systems, exhibiting a stable stress plateau, delayed densification, and substantial frictional dissipation during collective rearrangement. These results provide a mechanics-based route for designing shell assemblies with robust, direction-insensitive energy absorption.

**Keywords**: Eversion buckling, Shell stability, Post-eversion bistability, Energy absorption, Granular metamaterials

## 1. Introduction

Thin elastic shells are among the most efficient load-bearing components, offering high structural stiffness and load-carrying capacity through the transmission of external loads via in-plane membrane forces. The theoretical foundations for analyzing these structures were established by Love [1], who extended Kirchhoff's hypotheses to curved surfaces, and subsequently refined by Reissner [2] and Sanders [3] to account for the intrinsic coupling of membrane and bending energies—a duality that lies at the heart of shell stability. However, the real complexity of shell behavior is revealed in the nonlinear regime, where geometric nonlinearity and imperfection sensitivity often lead to a dramatic discrepancy between theoretical buckling predictions and experimental observations. As pioneered by von Kármán and Tsien [4] and later formalized by Koiter [5, 6], the initial post-buckling behavior of a shell determines its sensitivity to geometric defects, a principle that remains central to modern shell design methodologies.

A notable class of shell stability is the phenomenon of inversion, wherein a thin shell turns to assume an inverted configuration. For spherical caps, this inverted state approximates an isometry of the original surface, known as "mirror buckling", which in the limit of vanishing thickness would occur without elastic energy cost [7, 8]. The stability of such configurations depends critically on the energy landscape governed by the interplay between the shell's Gaussian curvature and its boundary conditions [7]. Analyzing these large-deformation transitions requires frameworks capable of handling finite rotations and large strains [9]. While classical shell theories are often formulated under shallow-shell assumptions, discrete differential geometry has emerged as a powerful numerical framework for treating shell inversion due to its nodal representation of geometry. Furthermore, analytical energy-scaling arguments also provide a general tool for assessing the stability by evaluating the competition between bending and stretching energies [10, 11].

Bistability arises when a shell admits two distinct equilibrium configurations separated by an energy barrier. For shallow shells, this behavior is controlled by a dimensionless parameter that measures the relative importance of stretching and bending [12, 13]. When this parameter exceeds a critical value, the inverted configuration can become stable [12, 14]. Otherwise, the shell relaxes to a single equilibrium state. Mechanically, this post-inversion stability is often determined by the stress state within the boundary layer [8, 15], which must accommodate the curvature change to maintain geometric continuity. If the resulting in-plane membrane stresses are sufficiently large, they drive a “snap-back” that restores the shell to its original state, thereby precluding bistability [16]. This energy-competition perspective is central to understanding multistability in systems ranging from biological structures to corrugated metasheets [13, 17].

Bistable shells have been widely used in energy-absorbing structures. Their performance relies on snap-through instabilities that dissipate energy through structural reconfiguration [18, 19]. Such systems also exhibit a relatively long and nearly flat stress plateau, a key characteristic of desirable protective materials [20, 21]. The energy dissipation in these architected metamaterials involves a complex interplay between unit-scale instabilities and system-scale phenomena, such as collective rearrangement and frictional sliding [22, 23].

Despite these advantages, most shell-based multistable absorbers suffer from inherent anisotropy. Because snap-through in conventional designs—such as von Mises trusses, arched shells, and asymmetrically laminated structures [24]—is typically directionally biased, these systems usually realize their full energy-absorption capacity only under appropriately aligned loads [25]. Under off-axis loading, these systems may fail to trigger the intended instability transition or exhibit severely degraded performance. This directional sensitivity constitutes a significant weakness in practical applications, where impact directions are often unpredictable and rarely aligned with the principal axes of the structure. Developing a mechanism that ensures robust bistability and energy dissipation regardless of loading orientation remains a significant challenge in mechanical metamaterials.

Here, we identify a new instability in everted open toroidal shells, which we term eversion buckling. The shell is generated by revolving a meridional arc about an external axis, producing a surface that is closed circumferentially but bounded by two circumferential free edges located at the meridional boundaries. Unlike a complete closed torus, these free edges permit the shell to be turned inside out. We show that the stability of the axisymmetric everted configuration after unloading is governed by a dimensionless parameter arising from energy scaling. Above a critical value, the axisymmetric state loses stability through a pitchfork-like bifurcation, leading to symmetry-breaking collapse. We demonstrate that this behavior can be tailored to achieve robust and repeatable bistability from any direction within the operational plane.

We combine theoretical analysis, finite element simulations, and experiments to characterize this behavior. The resulting instability is inherently isotropic due to the underlying axisymmetry. Building on this mechanism, we construct granular assemblies that exhibit stable stress plateaus and high energy absorption efficiency. These results establish eversion as a promising route to designing isotropic mechanical metamaterials.

## 2. Eversion buckling phenomenon

### 2.1 Geometry and eversion

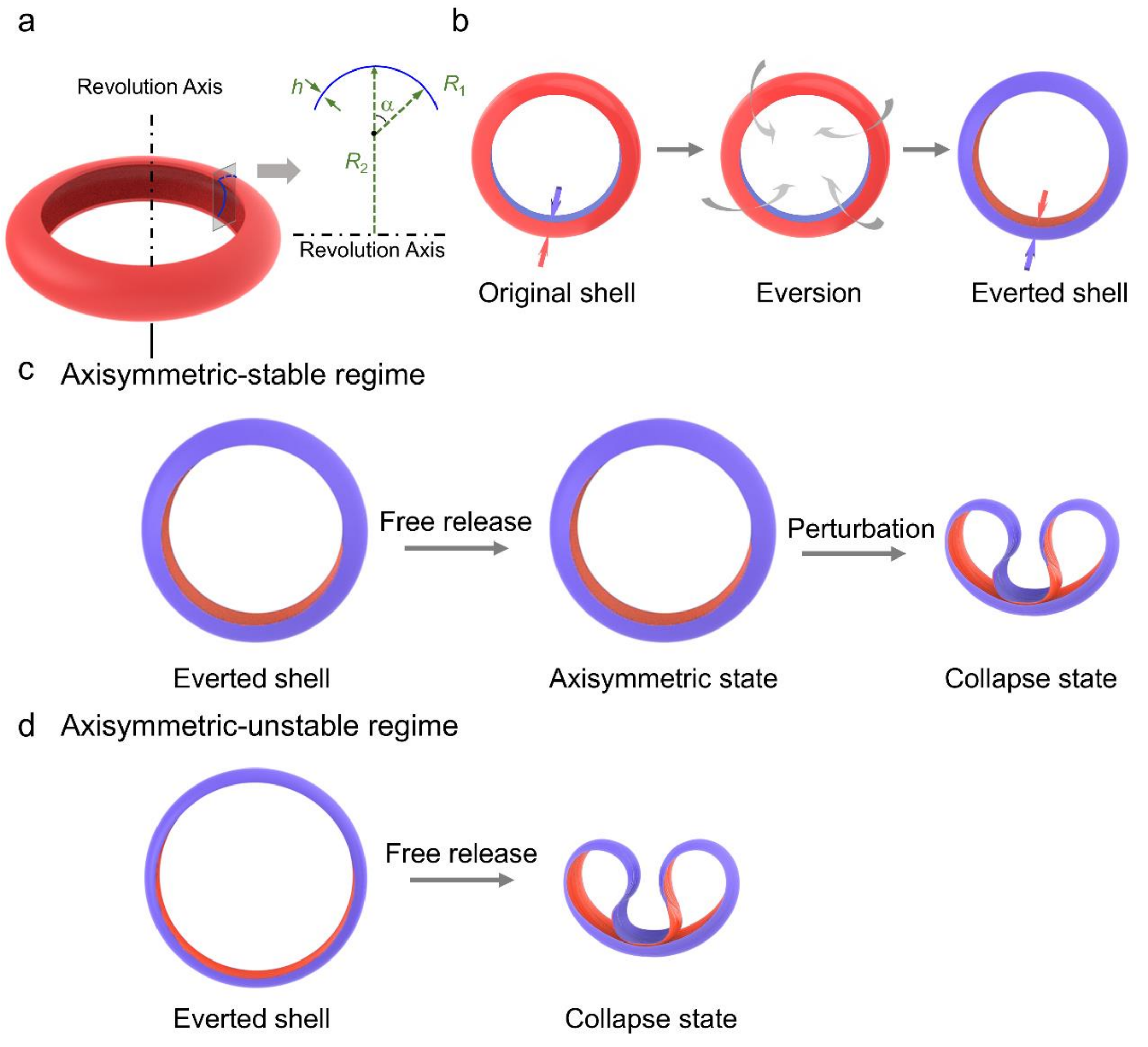

**Figure 1**. Eversion and post-eversion responses of an open toroidal shell. **a** Shell geometry. **b** Eversion process. **c** Load-free equilibria in the axisymmetric-stable regime:

the axisymmetric everted configuration and the symmetry-broken collapsed configuration reached after a finite perturbation. **d** In the axisymmetric-unstable regime, the axisymmetric everted configuration loses stability and collapses spontaneously after release.

We first develop the scaling analysis for a circular generating arc, whose constant meridional curvature permits a transparent derivation. The study is not restricted to circular arcs: semi-elliptical generating curves with nonuniform meridional curvature are subsequently examined and used in the experiments and granular assemblies. We consider a thin open toroidal shell generated by revolving a circular arc of radius $R_1$, at a distance $R_2$ from the axis of revolution (**Figure 1a**). The shell has uniform thickness $h$.

Eversion is imposed by forcing the shell to reverse its curvature, *i.e.*, turning inside out (**Figure 1b**). This process changes the sign of the principal curvatures while preserving the overall geometry of the midsurface, inducing a state of intense internal bending moments. Upon the removal of external constraints, the shell seeks an equilibrium configuration that minimizes the total stored elastic energy. The resulting self-stressed everted configuration stores elastic energy that can drive the subsequent morphological evolution of the shell.

**2.2 Post-eversion stability and eversion buckling**

After the shell is forced into an everted configuration and the external constraints are removed, the shell may follow different relaxation paths depending on geometry. The central question here is whether the axisymmetric everted configuration can remain as a stable equilibrium after unloading, or whether it loses stability and evolves into a symmetry-broken collapsed configuration. The classification used below is therefore based on the stability of the axisymmetric post-eversion equilibrium branch, rather than on a global count of every load-free equilibrium of the shell [26-29].

In the axisymmetric-stable regime, the shell retains its axisymmetric everted configuration after unloading (**Figure 1c**). This configuration is a local equilibrium that is stable against sufficiently small perturbations. A finite perturbation can nevertheless drive the shell toward a symmetry-broken collapsed configuration, which remains stable after the perturbation is removed (**Figure 1c**). Thus, two mechanically accessible load-free equilibria exist within the post-eversion state space in this regime: the axisymmetric everted configuration and the collapsed post-eversion configuration. This behavior may be described as post-eversion bistability.

In the axisymmetric-unstable regime, the axisymmetric everted configuration cannot persist after the eversion constraints are removed. The shell spontaneously breaks axisymmetry and evolves toward the collapsed configuration (**Figure 1d**), a transition referred to here as “eversion buckling”.

In all experiments and simulations considered here, unloading led either to a stable axisymmetric everted configuration or to a stable collapsed configuration, and spontaneous recovery to the natural uneverted state was not observed. Such recovery would first require a local region of the everted shell to be perturbed back toward its natural curvature, followed by the propagation of the recovered region over the shell. This recovery pathway is qualitatively different from the snap-back of a disk-like spherical cap, for which the boundary between inverted and natural regions can continuously contract and disappear at a point or reach the free edge. In the annular shell, by contrast, a locally recovered segment along the closed circumferential direction is bounded by a pair of transition zones separating regions with opposite curvature orientations. Complete uneversion therefore requires these transition zones to propagate around the annulus and eventually merge and disappear, or an equivalent large deformation involving the meridional free edges. The associated curvature reorientation and geometric compatibility can introduce additional localized bending, twisting, and membrane deformation, making complete recovery less readily accessible under the conditions examined here. Accordingly, the following analysis focuses on the two post-eversion configurations observed within the present experimental and numerical parameter range.

## 3. Energy scaling analysis of eversion buckling

We develop a scaling analysis to quantify the energetic competition in the everted shell. The shell is modeled using geometrically nonlinear, linearly elastic shell theory. Large rotations are allowed, while in-plane strains remain small. In addition, the shell is assumed to be thin and slender, with $h \ll R_1 \ll R_2$. Normals remain straight and normal to the midsurface, and transverse shear strains are negligible. The everted configuration is assumed axisymmetric; torsion and in-surface shear components vanish.

### 3.1 Kinematics

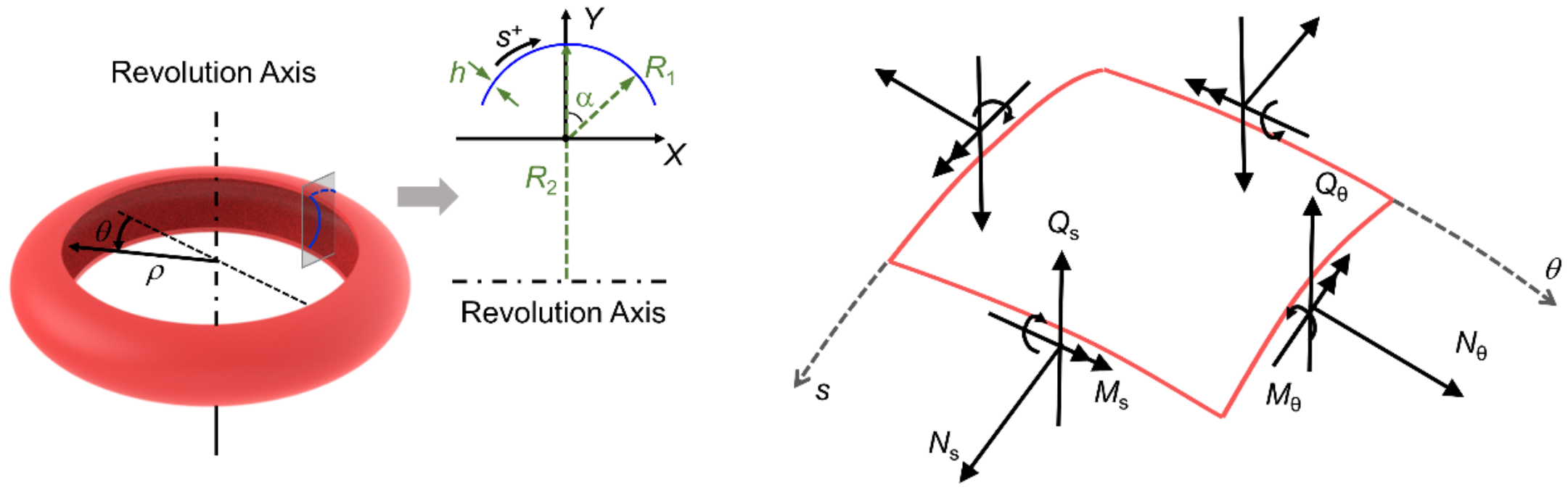

**Figure 2**. Schematic of the everted shell with a circular cross-section and the internal force and moment resultants.

The reference midsurface is denoted by $S_0$. A right-handed Cartesian coordinate system (*X*, *Y*, *Z*) is illustrated in **Figure 2** and set as follows. First, the generating circular arc of the shell, with a radius of curvature $R_1$, lies in the *XY*-plane. Second, the axis of revolution is a straight line parallel to the chord of the arc and located at a distance $R_2$ from the arc apex. This axis is denoted by $Y = Y_0,\ Z = 0$ with $Y_0 = R_1 - R_2$. Thus the *YZ*-plane is the plane orthogonal to the axis of revolution, and the radial direction in this plane will be parameterized by a circumferential angle $\theta \in [0, 2\pi)$. The generating circular arc is then parameterized in the *XY*-plane by the central angle $\alpha$ as follows:

$$X(\alpha) = R_1 \sin\alpha,\ Y(\alpha) = R_1 \cos\alpha,\ \alpha \in [-\alpha_{\max}, \alpha_{\max}], \tag{1}$$

in which $\alpha_{\max}$ represents the maximum central angle of the arc. In this study, finite-opening circular arcs are considered. The arc apex (farthest point from the chord) is at $\alpha = 0$, *i.e.*, $(X, Y) = (0, R_1)$, and the chord of the arc is the line $Y = R_1 \cos\alpha_{\max}$. The distance in the *YZ*-plane from a point $(X(\alpha), Y(\alpha), Z = 0)$ on the arc to the axis $Y = Y_0$ is

$$\rho(\alpha)=Y(\alpha)-Y_0=R_2+R_1(\cos\alpha-1). \tag{2}$$

This $\rho(\alpha)$ is the radius in the orthogonal *YZ*-plane of the circle traced by revolving $(X(\alpha),Y(\alpha),0)$ about the axis $Y=Y_0$.

A point of the reference midsurface $S_0$ corresponding to parameter $(\alpha,\theta)$ is obtained by rotating the point $(X(\alpha),Y(\alpha),0)$ around the axis $Y=Y_0$ by angle $\theta$ in the *YZ*-plane. Its coordinates are:

$$\mathbf{X}(\alpha,\theta)=\begin{bmatrix} X(\alpha) \\ \rho(\alpha)\cos\theta \\ \rho(\alpha)\sin\theta \end{bmatrix}=\begin{bmatrix} R_1\sin\alpha \\ (R_2+R_1(\cos\alpha-1))\cos\theta \\ (R_2+R_1(\cos\alpha-1))\sin\theta \end{bmatrix}. \tag{3}$$

Introducing the meridional arclength $s$ as the parameter instead of the angle $\alpha$, the above equation can be rewritten as

$$\mathbf{X}(s,\theta)=\begin{bmatrix} R_1\sin(s/R_1) \\ \rho(s)\cos\theta \\ \rho(s)\sin\theta \end{bmatrix}, \tag{4}$$

in which $\rho(s)=R_2+R_1(\cos(s/R_1)-1)$ and $s=R_1\alpha$.

The covariant tangent vectors of the reference midsurface are:

$$\mathbf{a}_s=\frac{\partial\mathbf{X}}{\partial s}=\begin{bmatrix} \cos(s/R_1) \\ \rho'(s)\cos\theta \\ \rho'(s)\sin\theta \end{bmatrix} \tag{5}$$

and

$$\mathbf{a}_\theta=\frac{\partial\mathbf{X}}{\partial\theta}=\begin{bmatrix} 0 \\ -\rho(s)\sin\theta \\ \rho(s)\cos\theta \end{bmatrix} \tag{6}$$

Since $\mathbf{a}_s\cdot\mathbf{a}_\theta=0$, the coordinates $(s,\theta)$ are orthogonal. The corresponding metric scale factors are:

$$A_s=1,\ A_\theta=\rho(s). \tag{7}$$

The reference principal curvatures (with a fixed choice of reference normal) are

$$\kappa_s^0 = \frac{1}{R_1},\ \kappa_\theta^0 = \frac{\cos(s/R_1)}{\rho(s)}. \tag{8}$$

After eversion, we describe the current midsurface $S$ by an axisymmetric parametrization:

$$\mathbf{x}(s,\theta) = \begin{bmatrix} z(s) \\ r(s)\cos\theta \\ r(s)\sin\theta \end{bmatrix}, \tag{9}$$

in which $r(s)$ is the current radius of the shell midsurface relative to the axis of revolution, and $z(s)$ is the coordinate along $X$-axis. Hence, we have

$$r'(s) = \cos\varphi(s),\ z'(s) = \sin\varphi(s), \tag{10}$$

where $\varphi(s)$ is the slope angle of the current meridian (the angle between tangent and $r(s)$), and prime $'$ denotes differentiation with respect to $s$. Then, the principal curvatures of the current midsurface $S$ are:

$$\kappa_s = \varphi'(s),\ \kappa_\theta = \frac{\sin\varphi(s)}{r(s)}. \tag{11}$$

Therefore, eversion changes the sign of the curvature field. For the purpose of energy scaling, and under the assumption of near-ideal eversion, the curvature changes are estimated as:

$$\Delta\kappa_s = \kappa_s - \kappa_s^0 \sim 1/R_1,\ \Delta\kappa_\theta = \kappa_\theta - \kappa_\theta^0 \sim 1/R_2. \tag{12}$$

The slenderness of the shell (meaning $R_2 \gg R_1$) implies that $\Delta\kappa_s \gg \Delta\kappa_\theta$.

### 3.2 Equilibrium equations in the everted shell

In $(s,\theta)$ coordinates, the nonzero stress resultants are the membrane forces $(N_s,\ N_\theta)$, bending moments $(M_s,\ M_\theta)$, and the meridional transverse shear resultant $Q_s$, as illustrated in **Figure 2b**. For an isotropic linearly elastic material with Young's modulus *E*, Poisson ratio $\nu$, the membrane and bending stiffnesses are

$$B = \frac{Eh}{1-\nu^2},\ D = \frac{Eh^3}{12\left(1-\nu^2\right)}. \tag{13}$$

Under small strains with finite rotations, the resultant-strain relations take the classical form

$$\begin{aligned} N_s &= B\left(\varepsilon_s + \nu\varepsilon_\theta\right),\ N_\theta = B\left(\varepsilon_\theta + \nu\varepsilon_s\right) \\ M_s &= D\left(\Delta\kappa_s + \nu\Delta\kappa_\theta\right),\ M_\theta = D\left(\Delta\kappa_\theta + \nu\Delta\kappa_s\right) \end{aligned}. \tag{14}$$

Crucially, since the post-unloading question concerns whether the everted shape can be a self-equilibrated state, we write equilibrium directly in the current configuration. Under axisymmetry, no torsion, and no external loads, balance of forces and moments yields the standard system [30, 31].

(i) Meridional Force Equilibrium:

$$\frac{\mathrm{d}N_s}{\mathrm{d}s} + \frac{r'}{r}\left(N_s - N_\theta\right) + \kappa_s Q_s = 0. \tag{15}$$

(ii) Normal Force Equilibrium:

$$\frac{\mathrm{d}Q_s}{\mathrm{d}s} + \frac{r'}{r} Q_s + \kappa_s N_s + \kappa_\theta N_\theta = 0. \tag{16}$$

(iii) Meridional moment equilibrium:

$$\frac{\mathrm{d}M_s}{\mathrm{d}s} + \frac{r'}{r}\left(M_s - M_\theta\right) - Q_s = 0. \tag{17}$$

Eqs. (15)-(17) explicitly display the geometric coupling via $r'/r$ and the curvature coupling through $\kappa_s$, $\kappa_\theta$.

Let $s = s^{\pm}$ denote the two meridional boundaries (corresponding to $\alpha = \pm\alpha_{\max}$ in the reference configuration). For traction- and couple-free edges,

$$N_s\left(s^{\pm}\right)=0,\ M_s\left(s^{\pm}\right)=0,\ Q_s\left(s^{\pm}\right)=0\,. \tag{18}$$

These conditions emphasize that any membrane forces in the unloaded, everted configuration are self-equilibrated through geometry and bending-stretching coupling, rather than being imposed by external tractions.

### 3.3 Scaling analysis of energy after eversion

To quantify the competing energy contributions in the post-unloading state, we introduce the following characteristic geometric scales. First, the characteristic geometric length along the meridional generating curve is estimated as

$$L \sim 2R_1\alpha_{\max}\,, \tag{19}$$

up to a constant factor of order unity. This quantity is analogous to the lateral shell size used in analyses of spherical-cap bistability [27], although here $L$ is measured along the curved meridian rather than in the projected plane. In addition, for the finite-opening circular arcs considered here, $\alpha_{\max}$ is of order unity, and hence $L \sim R_1$. Second, for our toroidal shell, the characteristic circumferential radius in the current configuration is of order

$$r \sim R_2\,. \tag{20}$$

Third, the area of the shell midsurface is of order

$$A_m \sim 2\pi r L \sim R_2 L\,. \tag{21}$$

At the scaling level, the bending energy density is

$$u_b \sim D\left[\left(\Delta\kappa_s\right)^2+\left(\Delta\kappa_\theta\right)^2\right] \sim D\left(\frac{1}{R_1^2}+\frac{1}{R_2^2}\right), \tag{22}$$

The above equation can be further simplified due to $R_2 \gg R_1$,

$$u_b \sim \frac{D}{R_1^2}. \tag{23}$$

Integrating Eq. (23) over the midsurface yields the total bending energy:

$$U_b \sim u_b \cdot A_m \sim D\left(\frac{R_2 L}{R_1^2}\right). \tag{24}$$

Membrane forces arise to maintain equilibrium. Their magnitudes are estimated sequentially from shell equilibrium equations. First, from Eq. (17), the transverse shear scales as the meridional gradient of bending moments,

$$Q_s \sim \frac{M_s}{L} \sim \frac{D\Delta\kappa_s}{L} \sim \frac{D}{R_1 L}, \tag{25}$$

Note that $r'/r \sim 1/R_2 \ll 1/L$ is utilized in the above analysis.

Second, we rewrite Eq. (15) in order-of-magnitude form as

$$\frac{N_\theta}{R_2} \sim \frac{N_s}{L} + \frac{D}{R_1^2 L}. \tag{26}$$

To identify the leading-order scaling, we first invoke a dominant balance argument by assuming that the left-hand term is balanced by the first term in the right-hand side in the above equation. This yields the relation

$$N_\theta \sim \left(\frac{R_2}{L}\right) N_s. \tag{27}$$

Because $L \sim R_1$ and $R_2 \gg R_1$, this scaling indicates a strong anisotropy in the membrane stress resultants, with the circumferential membrane force dominating the meridional one ($N_\theta \gg N_s$).

Third, from the normal equilibrium Eq. (16), we balance the shear gradient with the curvature-membrane coupling:

$$\frac{dQ_s}{ds} \sim \kappa_s N_s + \kappa_\theta N_\theta . \tag{28}$$

Evaluating the magnitude of each term yields:

$$N_s \sim \left(\frac{R_1}{L}\right) Q_s \sim \frac{D}{L^2} \tag{29}$$

and

$$N_\theta \sim \left(\frac{R_2}{L}\right) N_s \sim \frac{DR_2}{L^3} . \tag{30}$$

Substituting Eq. (29)-(30) into Eq. (26), we find the three terms in Eq. (26) possess the same order of magnitude, confirming the self-consistency of the dominant-balance assumption.

Finally, because $N_\theta \gg N_s$, the membrane strain energy density mainly comes from the circumferential force:

$$u_m \sim \frac{N_\theta^2}{B} \sim \frac{1}{B}\left(\frac{DR_2}{L^3}\right)^2 . \tag{31}$$

The total membrane energy scales as:

$$U_m \sim u_m \cdot A_m \sim \frac{D^2}{B}\frac{R_2^3}{L^5} . \tag{32}$$

Then, the ratio of the total membrane energy to the total bending energy is obtained by dividing Eq. (32) by Eq. (24):

$$\frac{U_m}{U_b} \sim \left(\frac{D}{B}\right)\left(\frac{R_1^2 R_2^2}{L^6}\right). \tag{33}$$

Substituting Eq. (13) into the above equation yields:

$$\frac{U_m}{U_b} \sim \frac{h^2 R_1^2 R_2^2}{L^6} . \tag{34}$$

Utilizing $L \sim R_1$, we thus arrive at the natural dimensionless parameter for a circular

generating arc:

$$\eta \propto \left(\frac{U_m}{U_b}\right)^{\frac{1}{2}} \sim \left(\frac{h}{L}\right)\left(\frac{R_2}{L}\right) \sim \left(\frac{h}{R_1}\right)\left(\frac{R_2}{R_1}\right). \tag{35}$$

This parameter is determined up to a constant factor of order unity and combines the effects of relative thickness and shell slenderness.

This parameter emerges naturally as the square root of the ratio of the membrane energy penalty to the bending energy. When $\eta \ll 1$, bending effects dominate, favoring the possibility of a statically admissible unloaded equilibrium within the axisymmetric everted configuration. As $\eta$ increases, the cost of membrane energy becomes dominant and makes the axisymmetric everted shape energetically prohibitive. To minimize this rising membrane energy, the structure must seek alternative deformation modes (*e.g.*, spontaneous collapse). The parameter $\eta$ therefore serves as a scaling measure for the onset of eversion buckling.

The preceding result is obtained for a circular generating arc. To examine whether the same scaling framework remains applicable to shells with nonuniform meridional curvature, we next consider a semi-elliptical generating arc with semi-axes $a$ and $b$. However, the meridional curvature varies along the arc and there is no single constant curvature radius equivalent to $R_1$. We therefore retain $L$ as a curvature-independent measure of the overall meridional size when extending the circular-arc scaling framework to the semi-elliptical geometries. For the moderate aspect ratios considered here, the characteristic meridional length and curvature radius remain of the same order. Using Ramanujan's approximation, we have

$$L \sim (a+b). \tag{36}$$

Therefore, Eq. (35) becomes

$$\eta \propto \left(\frac{h}{L}\right)\left(\frac{R_2}{L}\right) \sim \frac{hR_2}{(a+b)^2}. \tag{37}$$

## 4. Validation of the critical condition for eversion buckling

### 4.1 Experiments and finite element analysis

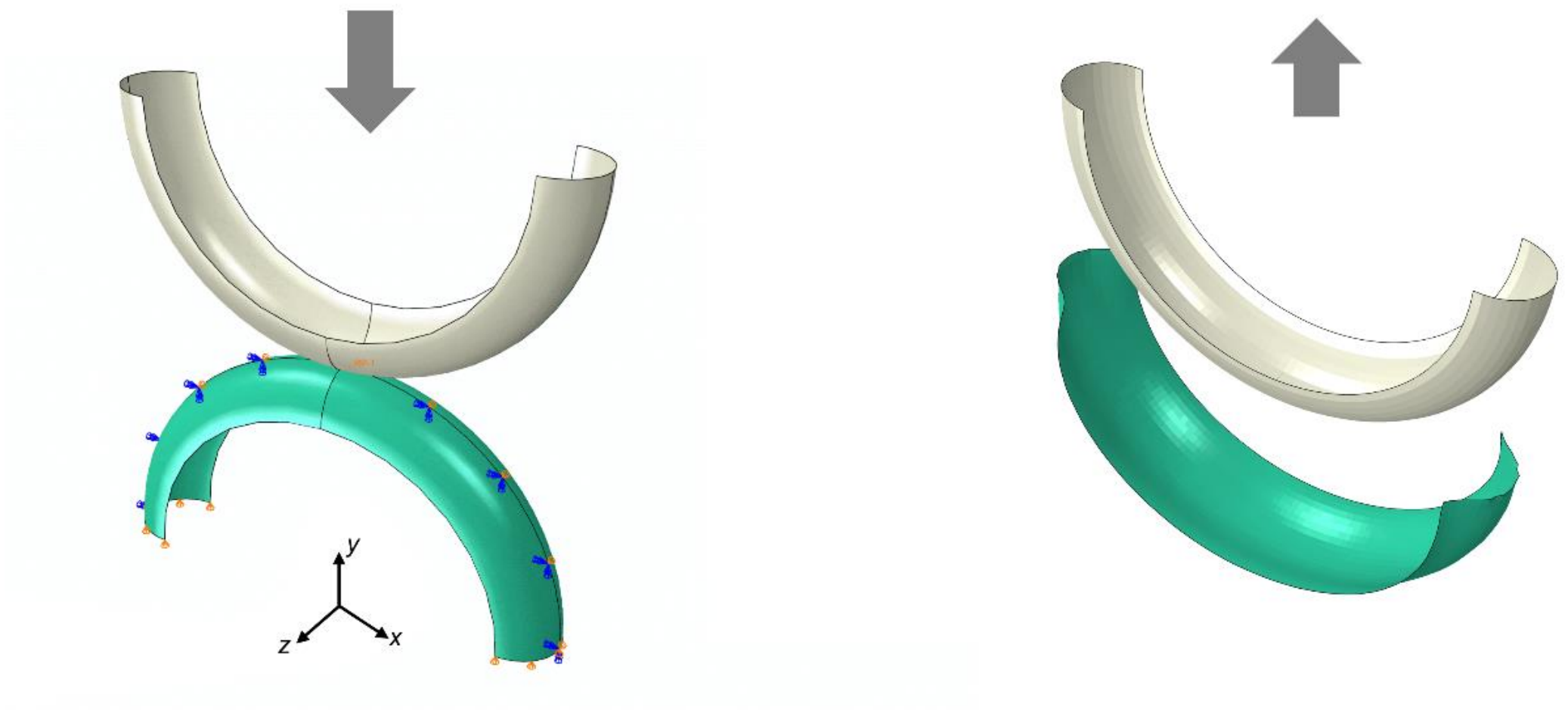

**Figure 3**. Finite element model for the eversion process of a half-shell geometry.

Toroidal-shell specimens with a uniform thickness of 0.5 mm were fabricated from thermoplastic polyurethane (HP-TPU) using Multi Jet Fusion on an HP Jet Fusion 5200 printer. The material was selected because of its high elastic limit and recoverable deformation, and its mechanical properties are provided in **Appendix A**.

To experimentally assess the post-eversion stability, each specimen was manually everted and then placed freely on a horizontal platform, after which all applied manipulation was removed. If the shell retained the axisymmetric everted configuration after unloading, it was classified as belonging to the axisymmetric-stable regime. If the axisymmetric configuration lost stability and spontaneously evolved into the collapsed state, it was classified as belonging to the axisymmetric-unstable regime. This procedure was repeated for shells with different geometric parameters.

The eversion process was also modeled using finite element analysis (FEA) in Abaqus/Explicit. Owing to the large deformation, rotation, and contact involved in the eversion of the complete open toroidal shell, a half-shell model is used to approximate the full toroidal geometry (**Figure 3**). The two cut edges of the half-shell midsurface are constrained to remain in the $xz$ plane by setting their $y$-displacement to zero. In addition, the major circumferential circle is constrained against $z$-displacement and rotation about the $x$- and $y$-axes, so that it remains in the same plane during eversion. Eversion is induced by pressing the half-shell against a rigid indenter having the same geometry as the shell to be everted, until the shell is fully turned inside out. The external load is then removed, and the system is allowed to relax for 5 seconds to reach the force-free equilibrium state after eversion. The numerical stability classification followed the same criterion as in the experiments: Retention of the axisymmetric everted configuration after unloading indicated the axisymmetric-stable regime, whereas spontaneous symmetry-breaking collapse indicated the axisymmetric-unstable regime.

The half-shell was modeled as Mooney–Rivlin hyperelastic solid, with the material parameters in **Appendix A**. The analysis employs four-node reduced-integration shell elements (S4R) with a mesh size of 0.6 mm, which is sufficient to capture boundary-layer gradients in bending moments and membrane forces for the geometry and shell thickness considered here (see **Appendix C**). Surface-to-surface contact was defined using hard contact in the normal direction and a friction coefficient of 0.2 in the tangential direction. Quasi-static conditions are ensured through controlled loading rates to minimize inertial effects while maintaining computational efficiency. Geometric nonlinearity is included to account for large rotations.

### 4.2 Membrane force distribution and stability phase diagram

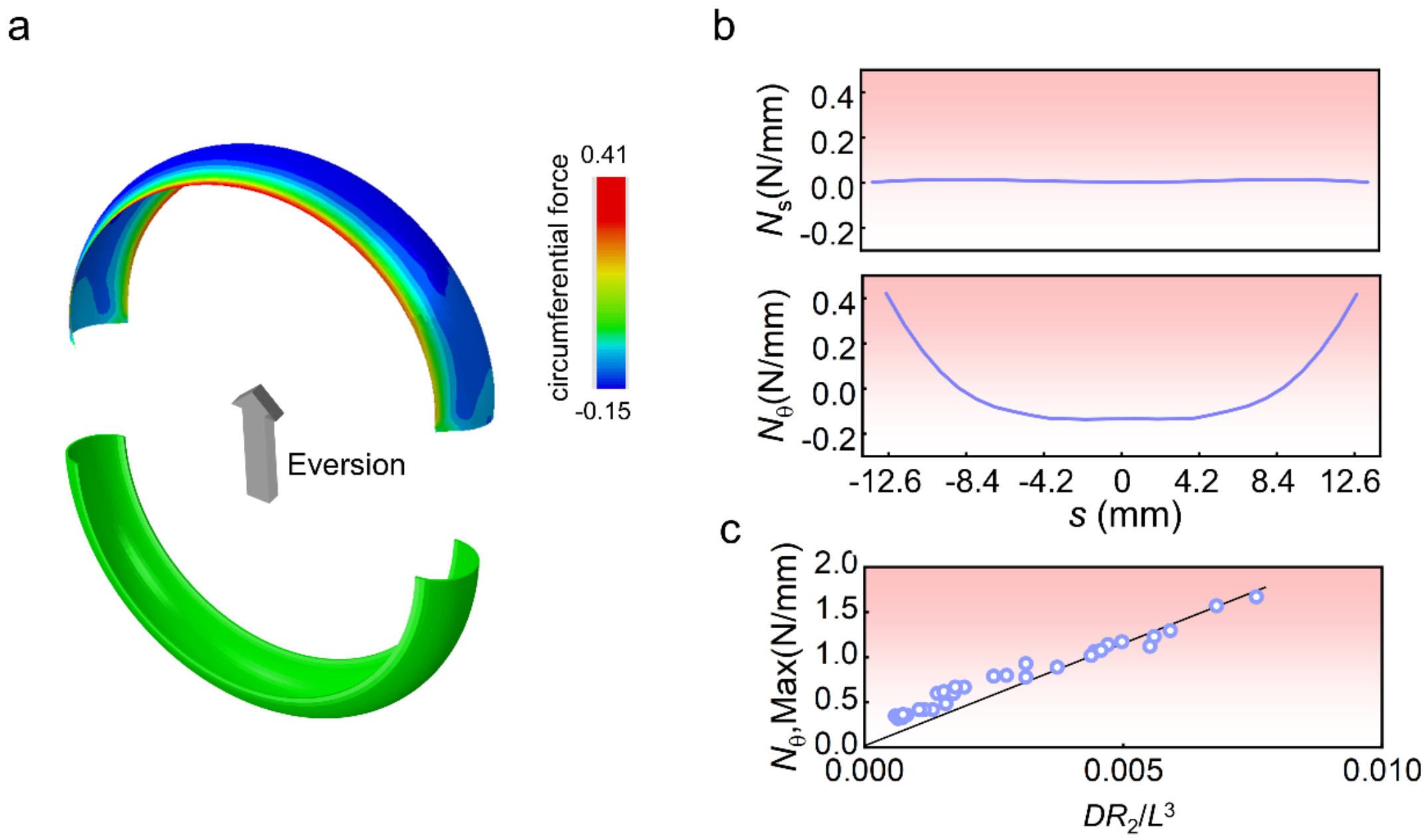

**Figure 4**. Membrane-force distribution and scaling in the everted half-shell. **a** Distribution of the circumferential membrane force $N_\theta$. **b** Meridional variation of the circumferential and meridional membrane forces. Shell parameters: semicircular meridian, $R_1 = 4$ mm and $R_2 = 20$ mm. **c** Maximum circumferential membrane force $N_{\theta,\max}$ as a function of $DR_2/L^3$ for different shell geometries. The solid line is a guide to the eye, indicating the linear trend of the FEA results.

The FEA results show that pronounced membrane forces are localized near the circumferential free edges (**Figure 4a**). Along the meridian, the circumferential membrane force ($N_\theta$) reaches its largest magnitude near the edges and becomes compressive in the central region (**Figure 4b**). Although this force field is spatially nonuniform, the scaling analysis characterizes its overall amplitude. Accordingly, the maximum absolute value $N_{\theta,\max}$ extracted from the simulations follows the same geometric dependence, exhibiting a linear relationship with $DR_2/L^3$ as given by Eq. (30) (**Figure 4c**). In addition, $N_\theta$ is substantially larger than the meridional force $N_s$ (**Figure 4b**). This strong anisotropy supports our theoretical scaling analysis (Eq. (30)): the energy penalty associated with the eversion process is primarily determined by the hoop stress required to maintain axisymmetry. The dominance of $N_\theta$ confirms that the bifurcation and stability of the everted state are governed by the circumferential membrane energy, justifying the use of the dimensionless parameter $\eta$ as a universal descriptor for the transition between post-eversion bistability and spontaneous collapse.

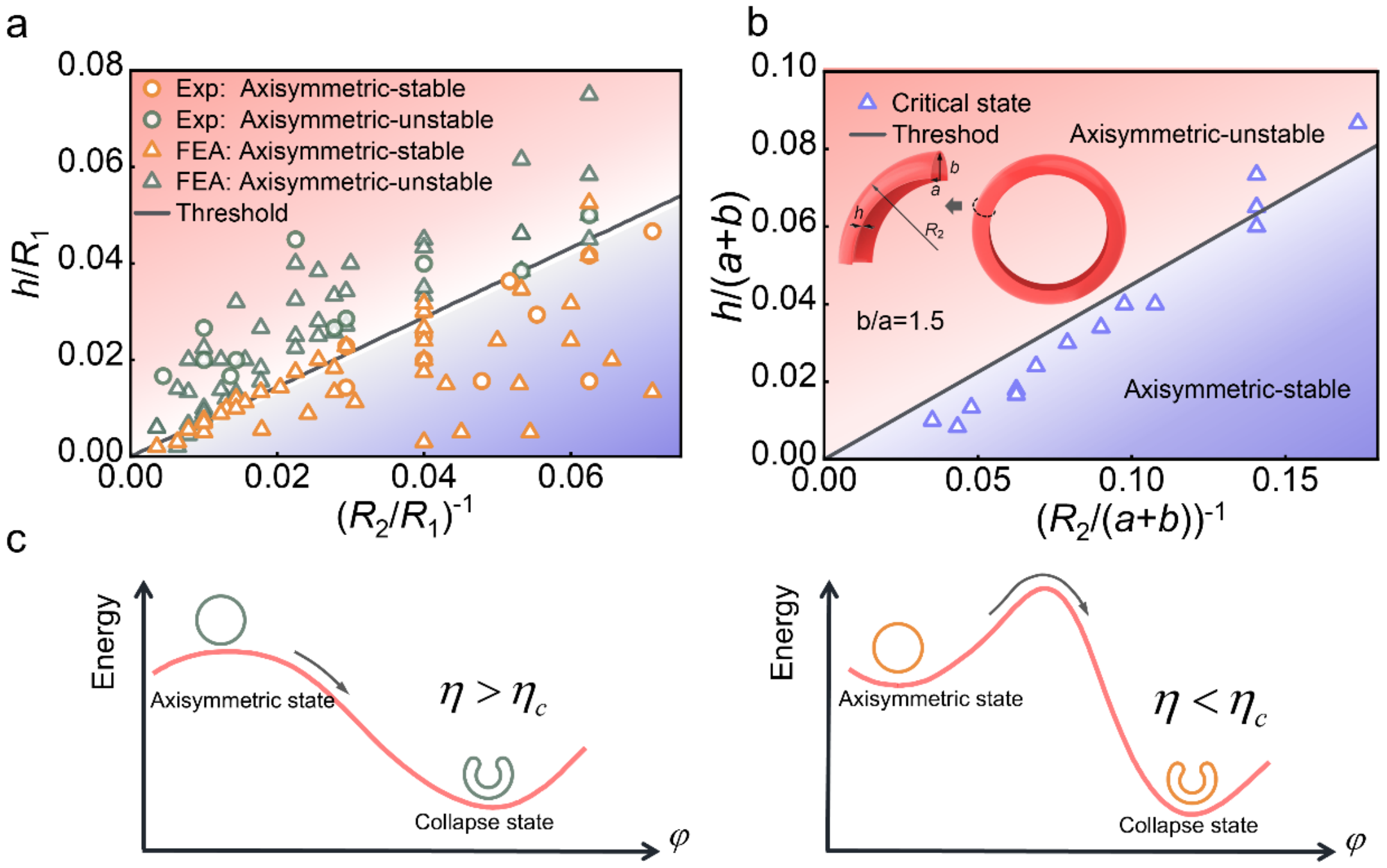

**Figure 5**. Stability phase diagram and scaling law. **a** Stability map for shells with semicircular generating curves. **b** Stability boundary for shells with semi-elliptical generating arcs. **c** Schematic post-eversion potential-energy landscapes expressed in terms of the generalized symmetry-breaking coordinate $\varphi$ for different values of the dimensionless control parameter $\eta$.

We constructed a stability phase diagram to identify the geometric boundary at which eversion buckling occurs after unloading. This boundary results from the competition between bending and membrane energies, characterized through $\eta$. By testing a wide range of geometric parameters via both experiments and FEA, we identify the limits of axisymmetric stability. For shells with a semicircular generating curve, **Figure 5a** displays that the data points collapse onto a single linear transition defined by $\left(R_2/R_1\right)^{-1} \propto \left(h/R_1\right)$ (*i.e.*, $\eta = \eta_c$, where $\eta_c$ is a material-independent constant for a given generating curve family). This confirms that the critical condition for eversion buckling is not merely a function of thickness alone but is intrinsically linked to the slenderness and the global curvature of the toroidal architecture.

Physically, this linear boundary represents an energy threshold where the energy penalty of maintaining an everted but axisymmetric shape exceeds the bending energy required to maintain the everted curvature. In the regime where $\eta < \eta_c$, the bending energy required for eversion is sufficient to balance the membrane energy penalty, allowing the everted axisymmetric shape to remain stable as a local energy minimum. Conversely, for $\eta > \eta_c$, the membrane energy becomes dominant, rendering the axisymmetric configuration energetically prohibitive and triggering a spontaneous transition toward a lower-energy collapsed state.

To examine whether the proposed scaling remains applicable when the meridional curvature is nonuniform, we further considered shells with semi-elliptical generating arcs. The corresponding FEA results are shown in **Figure 5b**. Despite the varying curvature along the meridian, the stability threshold remains linear. Importantly, the stability of these semi-elliptical shells can still be accurately predicted within the same dimensionless framework based on Eq. (37), supporting the applicability of the proposed scaling framework beyond circular generating curves. This result suggests that the eversion-induced instability is primarily governed by curvature and is not restricted to a specific cross-sectional geometry.

### 4.3 Energy landscape and the pitchfork-like bifurcation

The loss of stability of the axisymmetric everted branch can be interpreted schematically through the evolution of the system's potential-energy landscape (**Figure 5c**). Here, $\varphi$ represents a generalized order parameter characterizing the amplitude of the symmetry-breaking deformation. The axisymmetric everted state corresponds to $\varphi = 0$, and increasing $\varphi$ denotes an increasing departure from axisymmetry toward the collapsed configuration. Within the post-eversion pathway examined here, the collapsed configuration represents a lower-energy equilibrium than the metastable axisymmetric everted state.

Mathematically, this transition at $\eta = \eta_c$ corresponds to a pitchfork-like bifurcation. When $\eta < \eta_c$, the potential energy functional possesses a local minimum at the axisymmetric configuration, separated from lower-energy equilibrium by a potential barrier. In this state, the system exhibits bistability, requiring a finite external perturbation to initiate the snap-through. However, as the geometric parameter $\eta$ increases and crosses the critical threshold, the local potential well flattens and eventually transforms into a local maximum.

This bifurcation is fundamentally different from the saddle-node bifurcations typically observed in conventional bistable structures like curved beams or cylindrical caps [32, 33]. In those systems, stability is lost through the coalescence of a stable and an unstable equilibrium point. Instead, the present instability is a symmetry-breaking pitchfork-like bifurcation. Similar bifurcation structures have been reported in growing rods, shear-loaded pre-buckled bands, and geometrically constrained bigons [34-36], whereas here the instability arises from the self-equilibrated stress generated by shell eversion. Furthermore, the axisymmetry of the initial state ensures that the energy landscape is isotropic with respect to the azimuthal direction. Thus, at the critical point, the shell "falls" into the collapsed state by breaking this axisymmetry. This theoretical distinction is crucial: because the bifurcation is of the pitchfork type and originates from an axisymmetric state, the resulting snap-through is inherently omnidirectional within the equatorial plane. There is no predetermined "weak" direction for the collapse, a feature that we later exploit to design metamaterials with robust, direction-independent energy absorption capabilities.

## 5. Snap-through collapse from the stable axisymmetric everted state

### 5.1 Experiments and finite element analysis

To characterize the snap-through response of the post-eversion bistable shell, indentation tests were performed on toroidal shells. The specimens were fabricated from HP-TPU using Multi Jet Fusion, following the same manufacturing procedure described in Section 4.1. Before testing, each specimen was manually everted and allowed to relax into its stable axisymmetric everted state. A rigid indenter (hemisphere with radius of 2 mm) was then applied along the radial direction in the equatorial plane to trigger the transition to the collapsed state. The indentation tests were conducted using a displacement rate of 30 mm·min$^{-1}$. The force and indenter displacement were recorded continuously, and the deformation process was captured using a camera. For comparison, uneverted specimens were tested using the same indenter, loading direction, displacement rate, and support conditions.

The corresponding finite element analysis was conducted in Abaqus/Explicit. The shell was discretized using S4R elements with a characteristic element size of 0.6 mm. The TPU was modeled as a nearly incompressible Mooney–Rivlin hyperelastic material, with the parameters determined from uniaxial tensile tests as described in **Appendix A**. The indenter was modeled as an analytical rigid body, and surface-to-surface contact was defined using hard contact in the normal direction and a friction coefficient of 0.2 in the tangential direction. The same geometric dimensions, loading direction, and boundary conditions as in the experiments were used in the simulations. The support constraints were varied to investigate their effect on the mechanical response, while the shell geometry, prestress, indenter position, and loading history were kept unchanged.

Instead of directly simulating the complete eversion process, the residual-stress state of the everted shell was reproduced using an equivalent thermal-prestress method. The shell was represented by a three-layer composite section with layer-dependent thermal expansion coefficients. A uniform cooling step was first applied to generate the prestress, and the shell was then allowed to relax into its unloaded everted equilibrium before indentation. For comparison, the uneverted shell was modeled using the same geometry, material, contact, and loading conditions, but without applying the equivalent thermal prestress. The layer arrangement, thermal coefficients, and calibration procedure are provided in **Appendix B**. The model is therefore a calibrated surrogate for the post-eversion residual-stress state rather than a direct simulation of eversion. Quasi-static conditions before snap-through were checked by confirming that the kinetic energy remained below 2% of the internal energy, except during the short dynamic snap-through event.

### 5.2 Dynamic transition and energy release

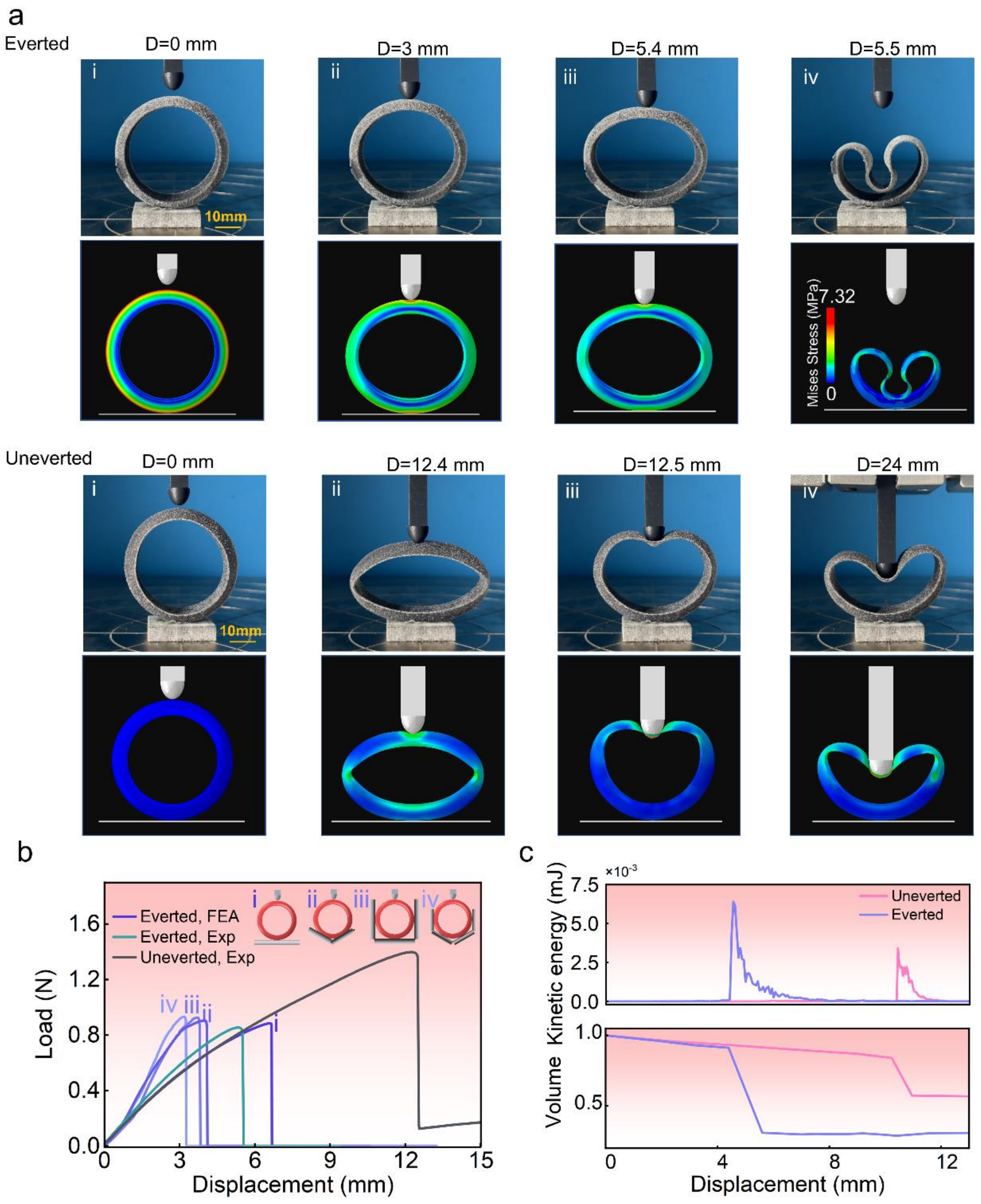

**Figure 6**. Dynamic transition and energy release of toroidal shells under indentation. **a** Experimental and finite element deformation sequences of the everted shell (top) and the uneverted shell (bottom). **b** Force–displacement responses of the everted shell under different support conditions, together with the experimental response of the uneverted shell. **c** Finite element predictions of the kinetic-energy evolution (top) and volume change (bottom) during the instabilities of the uneverted and everted shells. The geometry, support, contact, and loading conditions are identical to those used for the corresponding deformation sequences in **a**. The generating curve is semi-elliptical meridian with semi-axes $a = 3.5$ mm and $b = 5$ mm, while the shell thickness and major radius were $h = 0.5$ mm, and $R_2 = 20$ mm, respectively.

**Figure 6a** illustrates the dynamic transition of the everted shell from the initial axisymmetric state to the final collapsed configuration, in which experimental snapshots and FEA results show excellent agreement. The everted axisymmetric shell maintains its shape in a higher-energy equilibrium state due to the elastic energy stored during the eversion process. Unlike conventional bistable structures, such as shallow arches or von Mises trusses, where the snap-through is largely driven by the work of external forces, the collapse of an everted shell is primarily powered by the release of its internal strain energy. In this system, external loading acts only as a trigger to overcome the potential energy barrier associated with the pitchfork bifurcation. Correspondingly, the reaction force decreases rapidly toward zero after the critical point (**Figure 6b**), as the shell evolves into the lower-energy collapsed state.

For comparison, the shell without eversion-induced internal stress corresponds to the natural, uneverted configuration. Under indentation, it undergoes gradual deformation followed by localized buckling (**Figure 6a**). This instability also causes a decrease in the reaction force; however, the force remains nonzero and the shell remains in contact with the indenter (**Figure 6b**). The post-buckling configuration is therefore sustained by the applied load rather than representing a second stable equilibrium state. Thus, the uneverted shell may undergo a geometric instability, but it is not bistable.

A notable feature of the everted shell is that the critical force required to initiate snap-through varies only slightly among the different support conditions considered in

**Figure 6b**. This indicates that the onset of instability is governed mainly by the eversion-induced internal stress state rather than by the external constraints. Once the critical threshold is reached, the stored membrane and bending energies are rapidly converted into kinetic energy, driving the symmetry-breaking transition.

As shown in **Figure 6c**, the kinetic energy released by the everted shell is more than twice that of its uneverted counterpart. The transition is also accompanied by a pronounced volume contraction, whereas the uneverted shell exhibits only a limited volume change. The everted-shell transition occurs on a sub-millisecond timescale (~0.2 ms), combining rapid energy release with substantial volume reduction. These characteristics highlight the essential role of eversion in creating the bistability and large configurational change required for the energy-absorption mechanism developed below.

**5.3 Geometric mechanism of volume collapse**

A key result of this snap-through is the significant volumetric contraction, which reaches 61% (bottom panel in **Figure 6c**). This collapse occurs because the shell transforms from an axisymmetric everted shape into a non-axisymmetric folded structure.

In the everted state, the shell is forced into a configuration where its surface normal is opposite to its natural, stress-free state. When perturbed, the shell attempts to recover its original orientation to minimize its energy. However, due to the circular geometry and near-inextensibility, a kinematic incompatibility arises: the shell is unable to undergo a complete snap-back transition without engaging high-strain stretching modes.

To resolve this geometric conflict, the shell breaks its axisymmetry and localizes its deformation. Instead of a uniform reversal, it collapses into a folded state where the curvature is concentrated along narrow, hinge-like ridges (**Figure 6a**). This allows the opposing interior walls to move toward each other until they are nearly in contact, effectively closing the internal void. From a mechanics perspective, this folded geometry is a low-energy solution that satisfies the shell's curvature requirements through localized bending rather than global stretching. This specific transition is what leads to a volume loss three times greater than that of a shell without initial eversion.

### 5.4 In-plane omnidirectionality

The structural performance is further enhanced by the axisymmetry of the unit. In most bistable metamaterials, the snap-through is restricted to a single prescribed direction, such as the axis of a von Mises truss. In contrast, the everted toroidal shell can be triggered by a perturbation from any radial direction within the equatorial plane. Because the potential energy barrier is also axisymmetric, there is no predetermined "weak" direction for the collapse. As a result, perturbations applied at different radial angles lead to similar energy release and volumetric contraction.

This feature is critical for the collective behavior of the granular metamaterial discussed in the following section. It ensures that the assembly can respond to complex, multi-axial loading environments without a loss in energy-absorption efficiency. By utilizing the pre-stored energy of eversion and the geometric efficiency of the folded state, the everted shell in the axisymmetric-stable regime serves as a high-performance unit cell that connects discrete mechanisms with continuous shell theory.

## 6. Collective behavior: granular metamaterials for energy absorption

The transition from a single bistable unit to an assembled metamaterial introduces a complex interplay between local geometric compatibility and global dissipative mechanisms. By exploiting the omnidirectional snap-through of everted toroidal shells, we construct a granular metamaterial that leverages both discrete energy release and collective reconfiguration.

### 6.1 Assembly, mechanical testing, and simulation

To characterize the collective mechanical response, four assemblies were constructed using everted toroidal shells with uniform semi-elliptical generating curve ($a$ = 3.5 mm and $b$ = 5 mm) and thickness ($h$ = 0.5 mm). The corresponding major radii $R_2$ were 32.5 mm, 35 mm, 37.5 mm, and 40 mm, respectively. For each geometry, everted toroidal shells were arranged in a two-dimensional hexagonal packing and confined within a custom compression fixture (**Figure 7a**). The initial relative densities of the four assemblies were 0.09, 0.11, 0.12 and 0.15, respectively. The fixture consisted of two parallel glass plates with a length $L$ = 410 mm and a height $H$ = 370 mm, separated by $B$ = 15 mm to accommodate a single layer of shells with a thickness of approximately 13 mm. The glass plates restricted out-of-plane motion while allowing in-plane translation, rotation, and rearrangement of the shell units. The assembly was supported by a fixed bottom platen and compressed vertically by a rigid top platen. Quasi-static compression tests were performed using a Wance ETM254C testing machine equipped with a 100 N load cell at a constant platen speed of 10 mm·min$^{-1}$. The compressive force and platen displacement were recorded continuously.

A corresponding finite element model with the same packing arrangement and fixture dimensions was developed in Abaqus/Explicit (**Figure 7b**). Each shell was discretized using S4R shell elements with a characteristic mesh size of 0.6 mm and was assigned the same Mooney–Rivlin material model and equivalent prestress field as the single-shell model described in Section 5.1 and **Appendix B**. The top and bottom platens and the two confining plates were modeled as analytical rigid surfaces. The bottom platen was fixed, while the top platen was prescribed a vertical displacement. The confining plates restricted out-of-plane motion without constraining in-plane rearrangement.

Contact among the shells and between the shells and the fixture was modeled using the general-contact algorithm. Hard contact was used in the normal direction, and penalty friction with coefficients of 0.2 was used for shell–shell and shell–wall contact. The shells were initially placed in an ideal hexagonal arrangement. Gravity was applied first, and the assembly was allowed to settle for 3 seconds before compression. A mass scaling factor of 8 was adopted. During the quasi-static compression stage, the kinetic energy remained below 2% of the internal energy except during individual snap-through events, while the artificial strain energy remained below 1% of the internal energy.

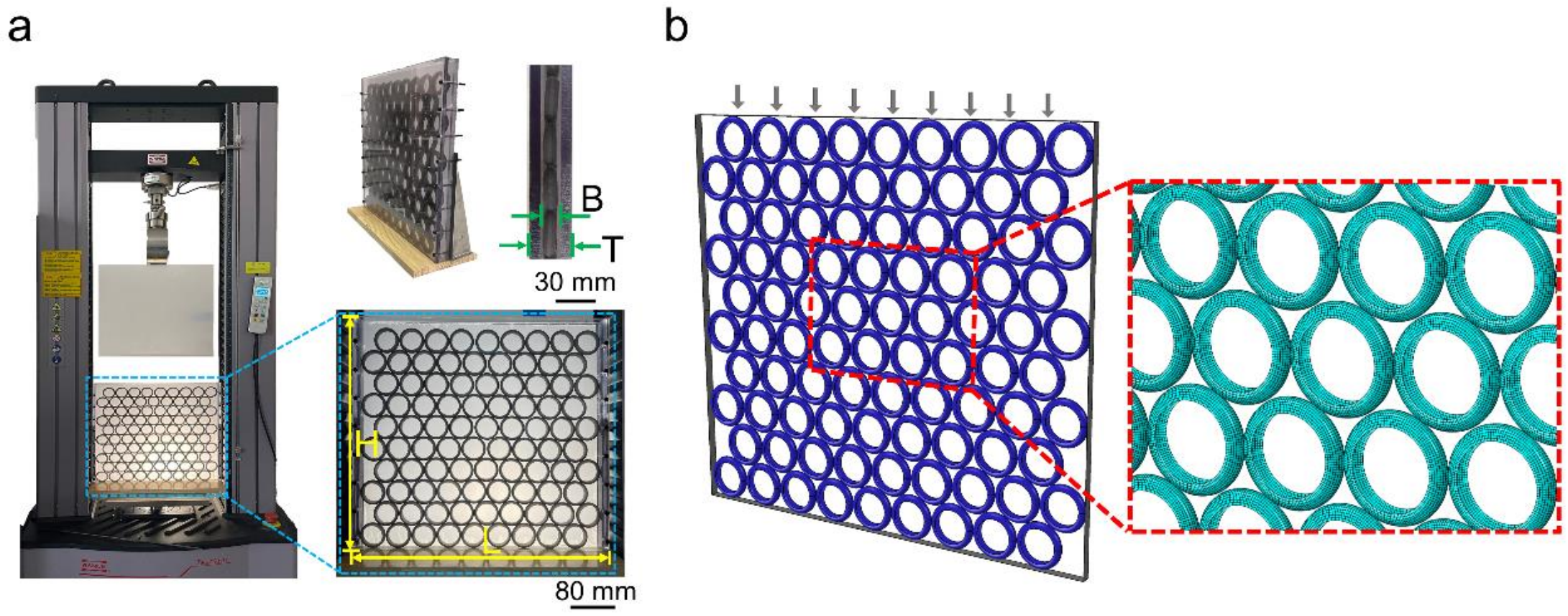

**Figure 7**. Experimental setup and FEA model of the metamaterial. **a** Uniform compression test setup. **b** FEA model and mesh.

### 6.2 Ideal energy absorption

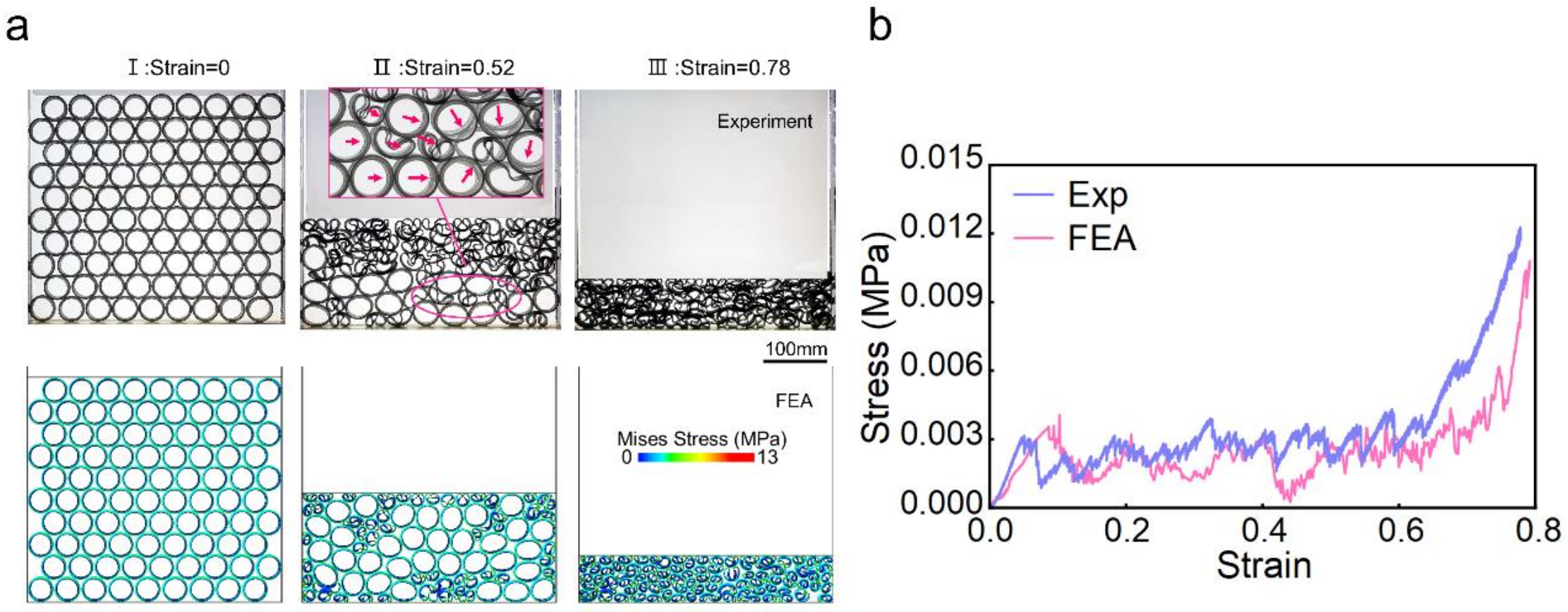

**Figure 8**. Progressive collapse of the granular metamaterial under quasi-static compression. **a** Experiments (top) and FEA (bottom). Inset: shell rearrangement into cavities created by collapsed neighbors. **b** Macroscopic stress – strain response. Shell parameters: $a = 3.5$ mm, $b = 5$ mm, and $R_2 = 40$ mm.

With increasing compression, the shells undergo a sequential snap-through collapse, leading to significant overall volume shrinkage (**Figure 8a**). The macroscopic compressive response of the assembled everted shells is characterized by a distinctive three-stage stress-strain trajectory (**Figure 8b**). Upon initial compression, the assembly shows linear elastic behavior driven by the contact between the curved boundaries of the everted units. As the stress reaches a critical threshold, the system enters a stable plateau regime. This plateau is caused by the sequential, symmetry-breaking snap-through of individual everted shells.

A fundamental mechanical advantage of this system lies in the axisymmetry of the everted state. In traditional cellular solids, the structural response is often sensitive to the loading angle relative to the cell walls. In contrast, the critical load of the everted shell is insensitive to boundary constraints (see Section 5.1), and its axisymmetric geometry allows for collapse from any radial direction. Consequently, the magnitude of the load drop caused by each individual collapse remains nearly constant, regardless of the shell's local contact orientation or the specific confinement within the assembly. This uniform load-drop behavior is the primary reason for the stability of the macroscopic stress plateau.

The most notable feature of this collective behavior is the delay in global densification, with strains exceeding 60% before densification occurs (**Figure 9a**). This delay is rooted in the large volume occupied by the everted shell. During snap-through, the shell transitions into a densely folded state, resulting in a massive volumetric contraction of approximately 61% at the unit-cell level. Furthermore, the independent shells provide a granular-like degree of freedom. Unlike fixed-node lattice structures, these shells can rearrange and move into cavities created by neighbors that have already collapsed (**Figure 8a** inset). This reconfiguration prevents early stress concentrations and contributes to the sustained plateau before densification. Consequently, the system achieves an exceptionally high Specific Energy Absorption Efficiency (SEAE, defined by the ratio of specific energy absorption to plateau stress of the assembly). As illustrated in **Figure 9b**, the SEAE of this assembly surpasses that of many conventional lattice materials, including those that rely on irreversible plastic deformation.

FEA further elucidates the energy evolution during compression of the granular assembly. Before and between successive snap-through events, deformation of the shells stores elastic strain energy. When individual shells collapse, part of this stored energy is released and drives local rearrangement, while sliding between neighboring shells produces accumulated frictional dissipation. The simulations show that the accumulated frictional dissipation exceeds the elastic strain energy remaining in the shells at the corresponding deformation stage (**Figure 10**). Together with the large deformation capacity and delayed densification enabled by sequential shell collapse, this substantial inter-shell friction contributes to the high energy-absorption performance of the assembly.

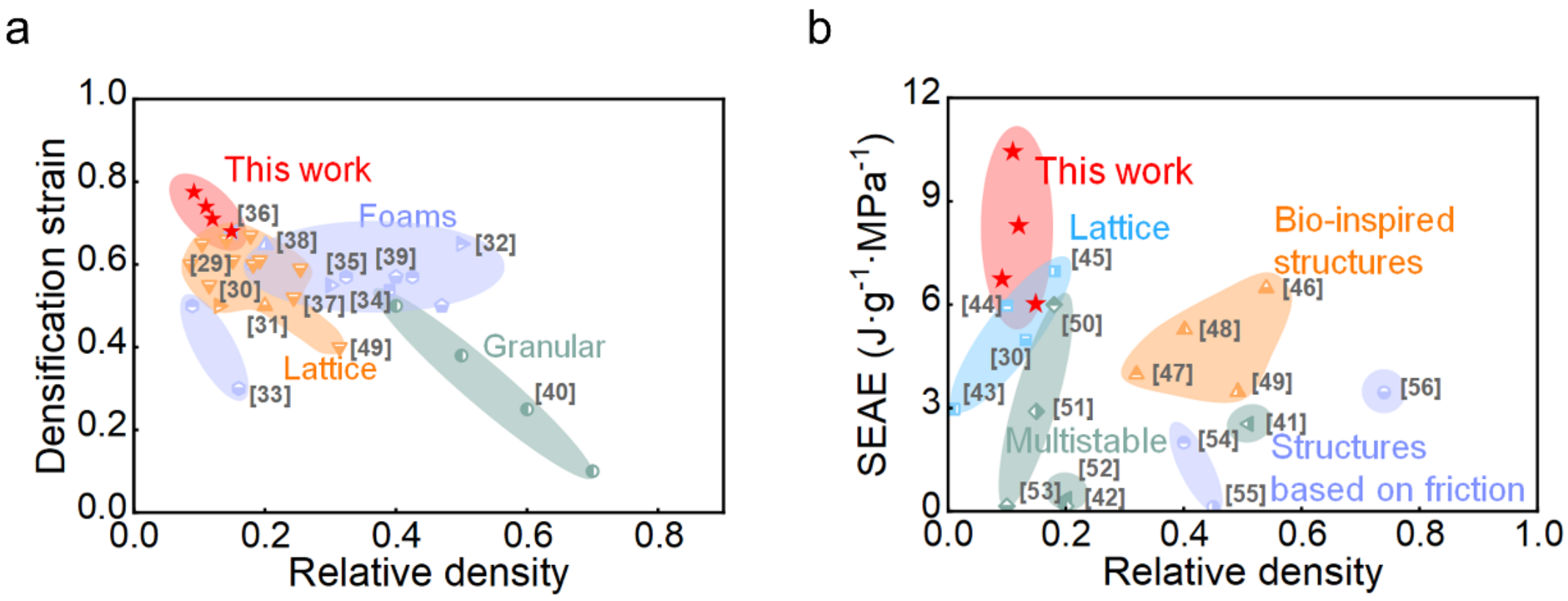

**Figure 9**. Ashby-type performance charts: **a** densification strain versus relative density [37-48]; **b** specific energy absorption efficiency versus relative density[38, 49-64].

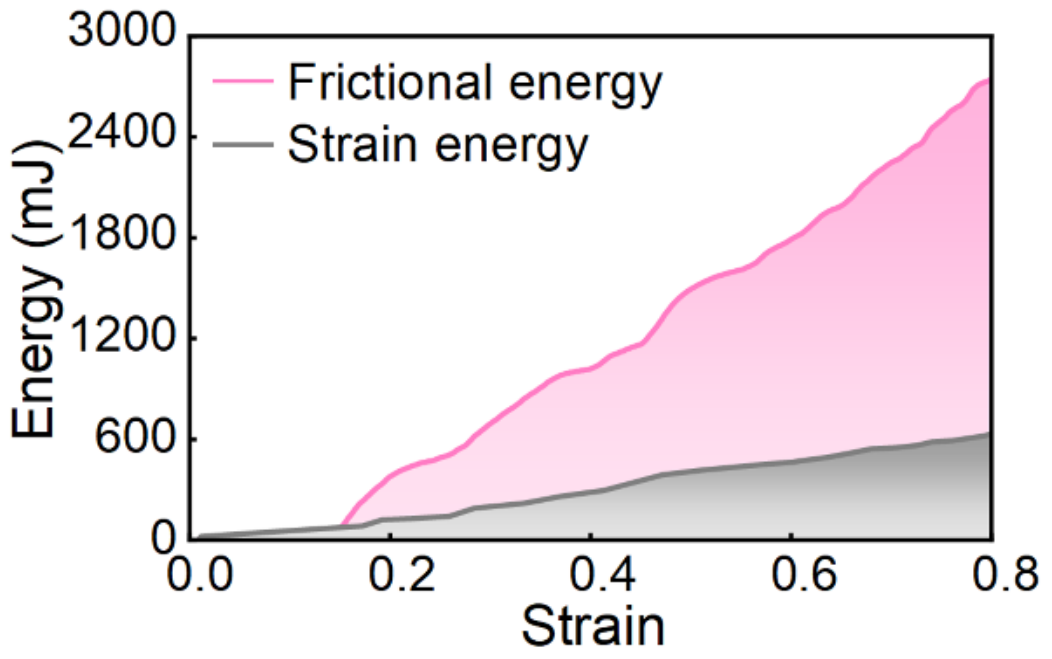

**Figure 10**. Evolution of the elastic strain energy stored in the shells and the accumulated frictional dissipation during compression, as predicted by FEA. Shell parameters: $a$ = 3.5 mm, $b$ = 5 mm, and $R_2$ = 40 mm.

### 6.3 Impact protection demonstration

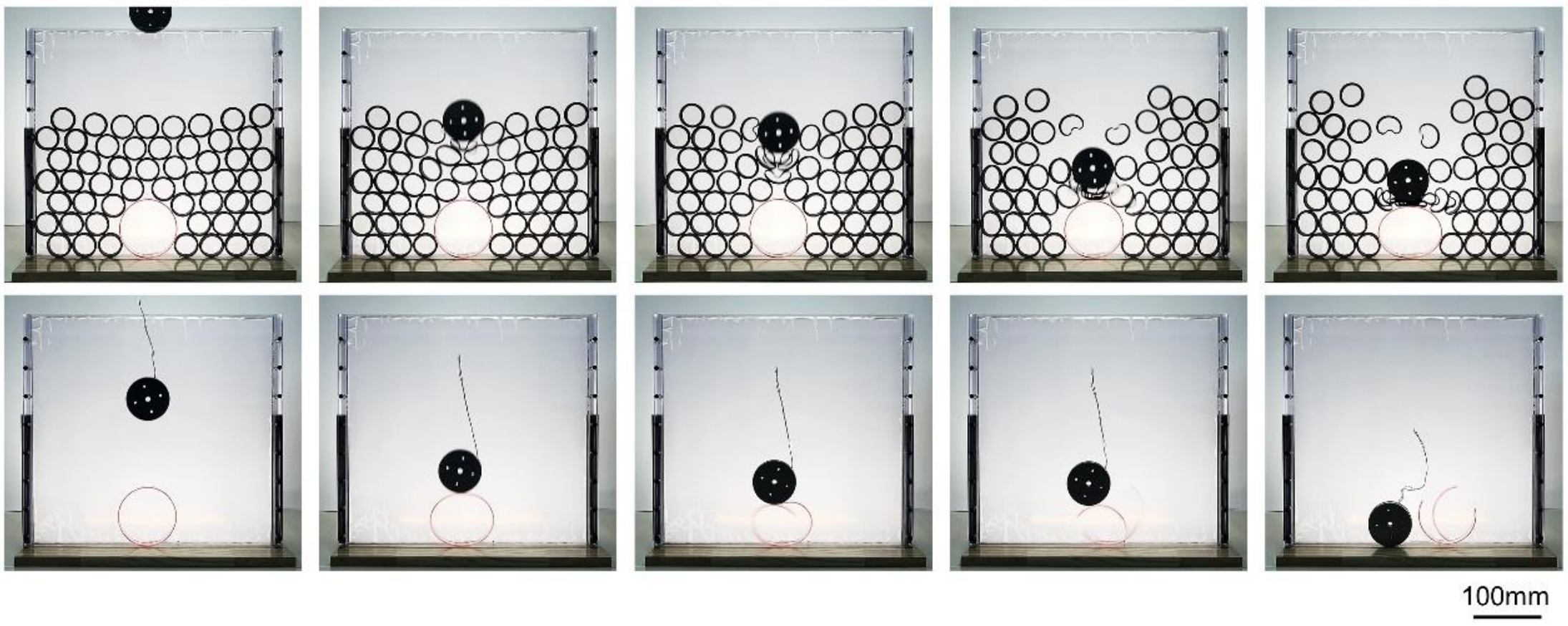

**Figure 11**. Impact-protection demonstration. A 0.42 kg metal disk was dropped from a height of 105 mm. Snapshots compare an unprotected fragile cylinder with one shielded by the everted-shell system. Shell parameters: $a$ = 3.5 mm, $b$ = 5 mm, and $R_2$ = 40 mm.

The practical utility of the everted-shell assembly is demonstrated through a proof-of-concept impact test. In a dynamic drop-test environment, the system acts as a high-performance mechanical filter. When a fragile PMMA surrogate is shielded by the assembly, the kinetic energy of the impactor is partitioned into the sequential, sub-millisecond snap-through events of the constituent shells. This collapse is enabled by the elastic energy pre-stored during eversion, allowing rapid snap-through, large deformation, and collective rearrangement under impact.

The system demonstrates a notable load-bearing capacity, effectively arresting the motion of impactors with a mass seven times greater than the assembly itself. This provides a robust, direction-independent protective mechanism that is rare in structured metamaterials. The combination of flowability, which allows the assembly to conform to complex object profiles, and the deterministic scaling of the energy release, positions these architectures as a versatile new class of impact-protective materials.

## 7. Conclusion

We have shown that the post-eversion stability of toroidal shells is governed by a competition between bending and membrane energies. This balance determines whether an everted shell remains in an axisymmetric metastable state or collapses into a non-axisymmetric configuration. A scaling analysis identifies a dimensionless geometric parameter that predicts the transition between the axisymmetric-stable and axisymmetric-unstable regimes, and this criterion is supported by finite element simulations and experiments across circular and semi-elliptical shell geometries.

The resulting instability, termed eversion buckling, is a symmetry-breaking transition. When the axisymmetric everted state loses stability, the shell collapses without a prescribed in-plane direction, consistent with a pitchfork-like bifurcation. In the axisymmetric-stable regime, eversion stores elastic energy that can be rapidly released through snap-through, producing large volumetric contraction and a triggering response that is weakly dependent on boundary constraints.

We further demonstrate that assemblies of everted shells in the axisymmetric-stable regime can function as granular energy-absorbing systems. Under compression, the shells collapse sequentially and rearrange collectively, leading to a stable stress plateau, delayed densification, and significant frictional dissipation. These results establish eversion buckling as a geometry-based mechanism for programming instability and energy absorption in thin-shell metamaterials.

## Acknowledgements

We acknowledge the support by the National Natural Science Foundation of China (Grants No. 12472071 and No. 12572170) and Beijing Key Fund Project (No. KZ20231000502).

**Appendix A: Constitutive modeling of thermoplastic polyurethane (TPU)**

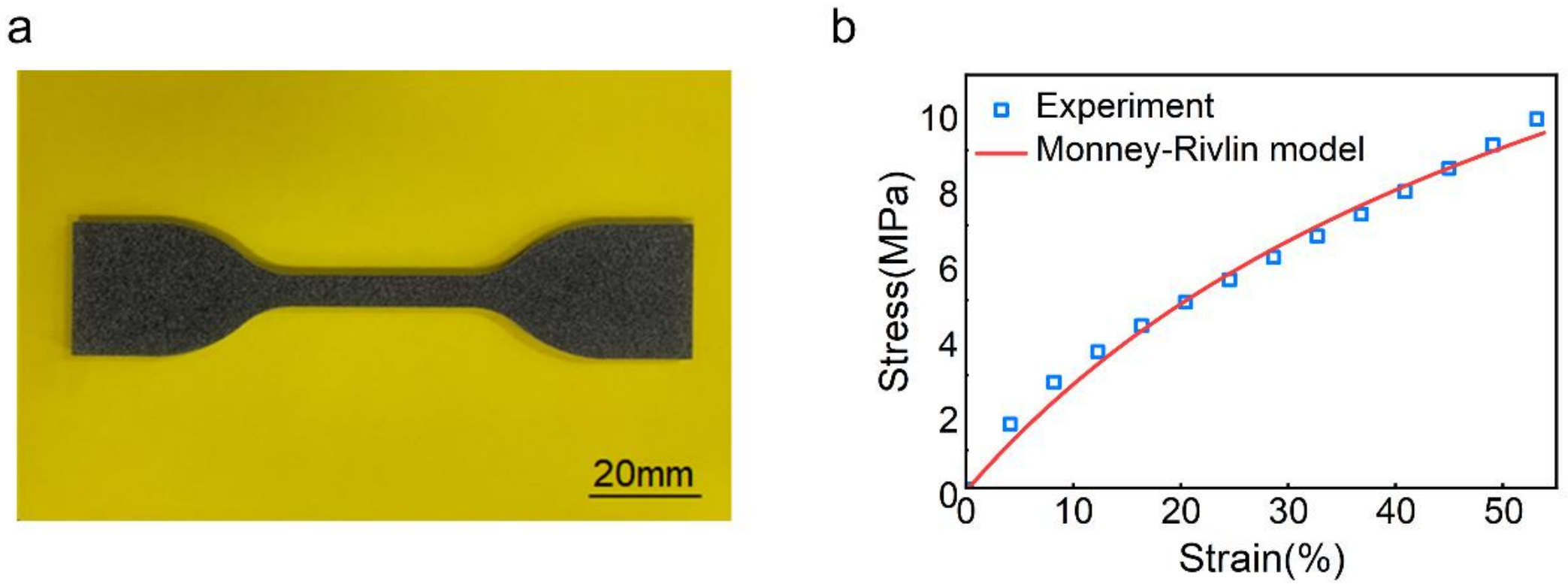

**Figure A1**. Mechanical properties of TPU. **a** Test specimen. **b** Stress–strain curve.

Thermoplastic polyurethane (TPU) samples were fabricated by 3D printing using HP Jet Fusion 5200. In finite element analysis, we use the Mooney-Rivlin hyperelastic model to describe the constitutive behavior of TPU, whose strain energy density is given by:

$$U = C_1(I_1 - 3) + C_2(I_2 - 3) \tag{A1}$$

where $I_1$ and $I_2$ are first and second deviatoric invariants of the right Cauchy–Green deformation tensor, $C_1$ and $C_2$ are material parameters. Under uniaxial tension, the corresponding Cauchy stress $\sigma$ can be written as

$$\sigma = 2\left(C_1 + \frac{C_2}{\lambda}\right)\left(\lambda - \lambda^{-2}\right) \tag{A2}$$

Where $\sigma$ and $\lambda$ are Cauchy stress and stretch in the tensile direction. Here, the material constants $C_1$ = 1.965MPa and $C_2$ = 3.513 MPa are fitted by Eq.(A2) against the experimental results (**Figure A1b**).

**Appendix B: Implementation of finite element analysis**

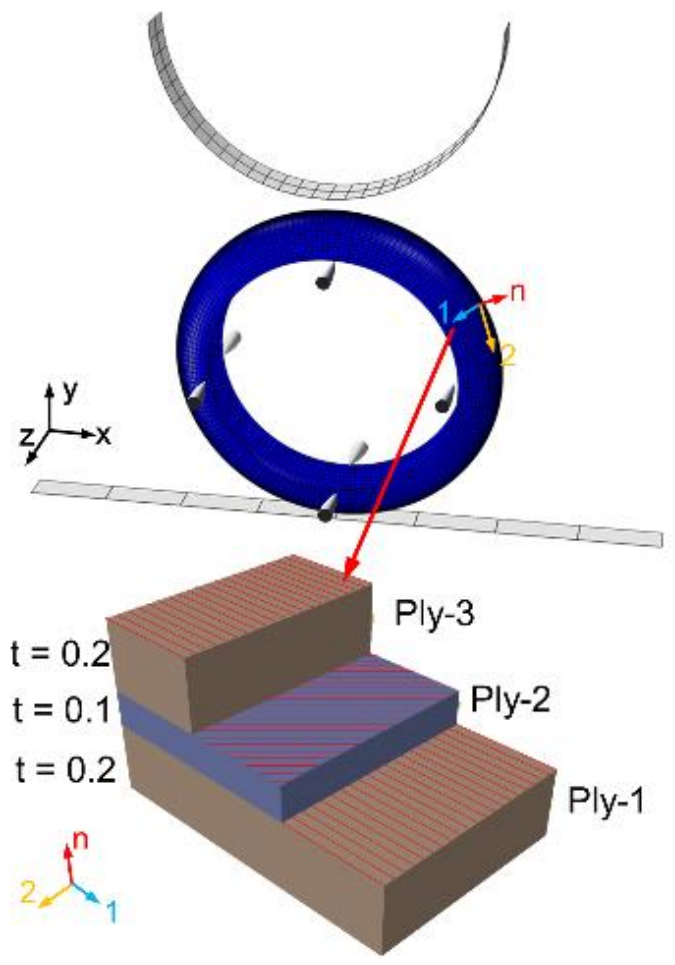

**Figure B1**. Finite element representations of the unit cell and toroidal shells. The unit cell is treated as a layered composite shell, with its everted configuration reproduced via a cooling-induced thermal loading that generates residual stress. The lower panel shows the three-ply stacking sequence (Ply-1 to Ply-3).

Direct simulation of full shell eversion in finite element analysis (FEA) is computationally prohibitive. Instead, a surrogate approach was adopted in which prestress is introduced via a thermal loading scheme; uniform cooling is used to approximate the residual stresses induced by eversion. The shell is modeled as a three-layer composite with layer thicknesses of 0.2, 0.1, and 0.2 mm, respectively, and layer-specific thermal expansion coefficients. A local coordinate system is defined (**Figure B1**): the 1-direction is tangent to the cross-section, the 2-direction is circumferential, and the n-direction is normal to the surface.

To reproduce the observed deformation, the coefficients (1-, 2-, n-directions) are assigned as follows: Ply-1 (−0.00035, −0.00023, 0 $K^{-1}$), Ply-2 (−0.0001, 0.0005, 0 $K^{-1}$), and Ply-3 (0.00035, 0.00023, 0 $K^{-1}$). Ply-2 is oriented at 45°, whereas Ply-1 and Ply-3

align with the 1-direction (**Figure 3a**). Prestress is generated through a dedicated thermal step with uniform cooling from 120 K to 0 K. The parameters are identified by iterative calibration against experiments, targeting agreement in critical load, post-collapse mode, and critical strain. The resulting model is therefore partially phenomenological.

## Appendix C: Mesh convergence

To assess the adequacy of the adopted mesh resolution, a mesh-convergence analysis was performed using nominal element sizes of 0.4, 0.6, and 0.8 mm, while all other geometric, material, boundary, and loading parameters were kept unchanged.

As shown in **Figure C1**, the three meshes produce essentially identical post-eversion configurations and membrane-force distributions, with no appreciable differences in the boundary-layer region. An element size of 0.6 mm was therefore adopted as a balance among numerical accuracy, computational cost, and the smoothness of the extracted membrane-force profiles.

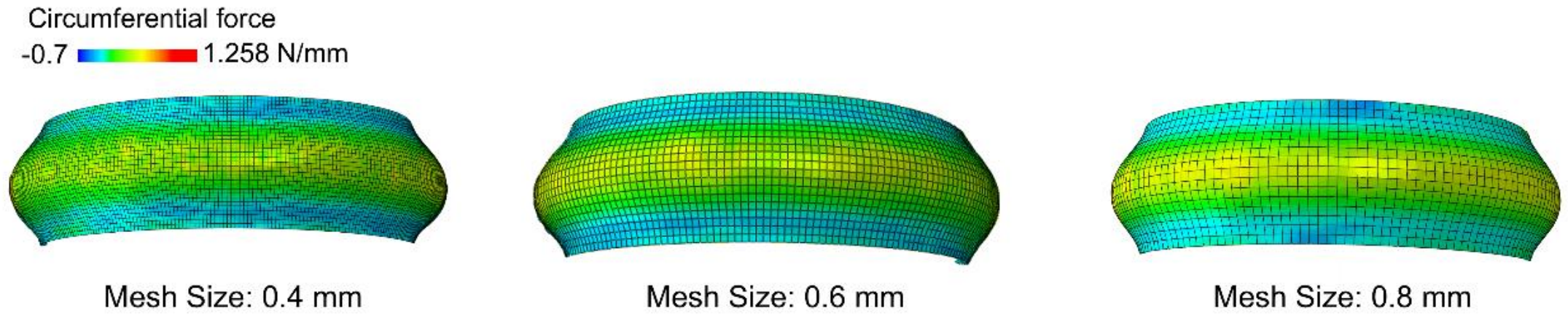

**Figure C1**. Mesh-convergence analysis of the post-eversion finite element model. Post-eversion configurations and corresponding membrane-force distributions with nominal element sizes of (a) 0.4 mm, (b) 0.6 mm, and (c) 0.8 mm. The shell thickness is $h$=0.5 mm, and all other model parameters are identical.